# Nurturing the Industrial Accelerator Technology Base in the US

A. M. M. Todd, AMMTodd Consulting

R. Agustsson, A. Murokh, M. Ruelas, RadiaBeam Technologies, LLC.

D. L. Bruhwiler, RadiaSoft LLC.

K. Yoshimura, Research Institute for Interdisciplinary Science, Okayama University

J. W. Rathke, J. W. Rathke Engineering Services

J. Chunguang, A. Kanareykin, Euclid TechLabs LLC.

S. C. Gottschalk, STI Magnetics

V. Yakovlev, Fermilab

FOREWORD

The purpose of this white paper is to discuss the importance of having a world class domestic industrial vendor base, capable of supporting the needs of the particle accelerator facilities, and the necessary steps to support and develop such a base in the United States. The paper focuses on economic, regulatory, and policy-driven barriers and hurdles, which presently limit the depth and scope of broader industrial participation in US accelerator facilities. It discusses the international competition landscape and proposes steps to improve the strength and vitality of US industry.

# 1 Table of Contents

## 2  Executive Summary

There is a widespread perception within US industry that to the United States Department of Energy (DOE), and to the DOE National Laboratories, a transfer of accelerator technology to US Industry is not a high priority. It is not uncommon for the US high energy physics community to develop state-of-the-art particle accelerator technology, and later having to buy that same technology from abroad for domestic projects, because US firms are neither supported nor encouraged to partake in the DOE programs of record.

In contrast, the high energy physics communities in Europe and Asia work to nurture their domestic industrial bases, and this asymmetry in technology transfer policy creates an uneven playing field when US firms attempt to compete overseas, while greatly benefits foreign companies competing to serve DOE funded projects in the US. The purchase of superconducting RF cavities for the LCLS-II upgrade is considered in some detail as a case study.

This resultant relative weakness of the accelerator technology industrial base in the US, has many undesirable consequences, including increased costs and reduced availability of critical components required by the labs, excessive and often dangerous US dependence on foreign sources, a geographically localized and socially narrowed recruitment base for the technical personnel involved in the accelerator projects, and reduced recognition by society of benefits associated with the government investments into accelerator science and technology.

The US Small Business Innovation Research (SBIR/STTR) program is a great asset to help small businesses develop new capabilities, and it is the envy of many other countries, but the DOE does little to nurture these small businesses across the "Valley of Death". Case studies presented discuss hardware and software projects and initiatives developed through the SBIR program.

The paper covers other relevant areas of the relationship of accelerator industry and National Laboratories, including:

- knowledge transfer needs and a role DOE could play in facilitating such efforts,
- limitations of the current technology transfer programs,
- industry-laboratory collaboration in recruitment and training of the necessary qualified personnel (especially engineers and technicians),
- how to mitigate the risks of prototype development projects,
- industry-laboratory collaboration to gain access to equipment and expertise, and
- a need to streamline and simplify the procurement process to make it more accessible and less burdensome to small businesses and professional shops, willing and capable of serving the needs of the DOE National Laboratories.

Many of the issues raised and recommendations issued are based on the author's experience and understanding of where this industry is today, and how to make it more viable over the coming decade. The HEP community motivation, political will, and legal framework to introduce such changes are outside the scope of this discussion.

## 3  Overview of the US accelerator facilities needs

Accelerator facility construction and upgrades in the US are funded through the DOE Office of Science, with an FY2022 budget request close to $7.5 billion. The combined budget of the Offices of Basic Energy Sciences (BES), High Energy Physics (HEP), and Nuclear Physics (NP), which are the main DOE offices supporting accelerator facilities, science, and technology, is about $4 billion (Fig. 1). Out of this amount, $603 million is requested in support of accelerator facility upgrade and construction projects. This amount excludes smaller scale R&D funding, and various activities in support of commercial applications (e.g. radiotherapy and cargo inspection).

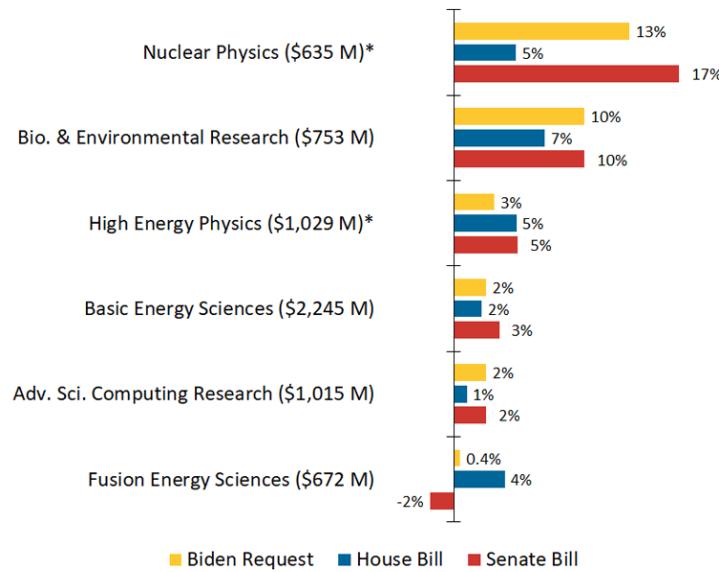

| Project | DOE component | FY22 budget request |
|---|---|---|
| Spallation Neutron Source (SNS) upgrade | BES | $17 M |
| SNS Second Target Station construction | BES | $32 M |
| Linac Coherent Light Source II (LCLS-II) construction | BES | $90 M |
| LCLS-II energy upgrade | BES | $53 M |
| Advanced Light Source (ALS) upgrade | BES | $75 M |
| Advanced Photon Source (APS) upgrade | BES | $106 M |
| National Synchrotron Light Source II beamlines | BES | $15 M |
| PIP-II upgrade at Fermilab | HEP | $90 M |
| Mu2E at Fermilab | HEP | $13 M |
| Facility for Rare Isotope Beams (FRIB) construction | NP | $82 M |
| Electron-Ion Collider (EIC) construction | NP | $30 M |
| *Total* | *DOE* | *$603 M* |

**Fig. 1:** (Top) FY22 DOE Office of Science budget increase proposals in %, with FY21 numbers listed in parentheses on the left (adapted from [1]); (bottom) new accelerator facility projects and upgrades in the US, and their requested FY22 budgets, from Ref. [1].

Thus, the US accelerator facility upgrades, and construction projects represent a significant addressable market for industry. They require various industrially produced components and sub-systems, including superconducting and normal conducting accelerating cavities, insertion devices, magnets, diagnostics and instrumentation, lasers, microwave sources and systems, power electronics, ultra-high vacuum systems, controls, and control electronics; as well as advanced scientific and engineering computational software, tools, and capabilities.

Companies around the world offer products and services that can meet these needs. The unusual fact, however, is that the participation of the US domestic industry in providing accelerator technology products to these facilities and projects has been uncharacteristically minimal. The leadership and dynamism, which generally are trademark features of US business and industrial culture elsewhere, for some reasons have not been applied towards serving the needs of the accelerator facilities, and when applied – not sustained for too long.

This white paper is an attempt to reflect on that observation. We first discuss why the industry role is important, and then try to analyze why industrial participation is currently lacking and what can be done to improve this in the future.

# 4  Why domestic industrial participation is important

As recent events around the world have highlighted, securing domestic supply chains for the US-based accelerator complex is necessary in order to minimize operational and construction risks and components shortages. Cooling relations with China and sanctions against Russia have affected the US supply chains for accelerator technology. It is pertinent to note that companies and institutes within these two countries have routinely served as discounted suppliers for many critical accelerator components. This is a development long in the making, built on decades of collaboration of Chinese and Russian institutions with U.S. technical staff to gain the knowledge and skills necessary to sell products and systems to the US accelerator facilities and light sources.

All this time invested in foreign entities cannot be easily transferred back to domestic industry and represents 'lost' investments when the political climate changes. In the interim, the once viable domestic vendors have either a) folded due to the inability to compete with state-subsidized foreign vendors, or b) exited the market to focus on lower-risk, higher-margin products. This results in fewer competitors both globally and domestically, which in turn results in lower quality, longer lead times, and higher-cost products. Such overdependence on foreign suppliers can easily become detrimental to domestic R&D programs. For example, at the time of this writing, the lead time for a superconducting accelerating cavity ordered from Europe is over 18 months and growing, while there are no longer viable domestic alternatives.

Besides the supply chain issues, there are also national security implications when a high-tech domestic industrial sector does not attract sufficient support to be sustainable and is eventually overtaken by Chinese suppliers. China is a very capable and proactive global competitor to the US, and every time another relevant industrial process loses its commercial viability domestically, this eventually directly contributes to the success of the "Made in China 2025" program, which is already on par with the US in many critical technologies (Fig. 2).

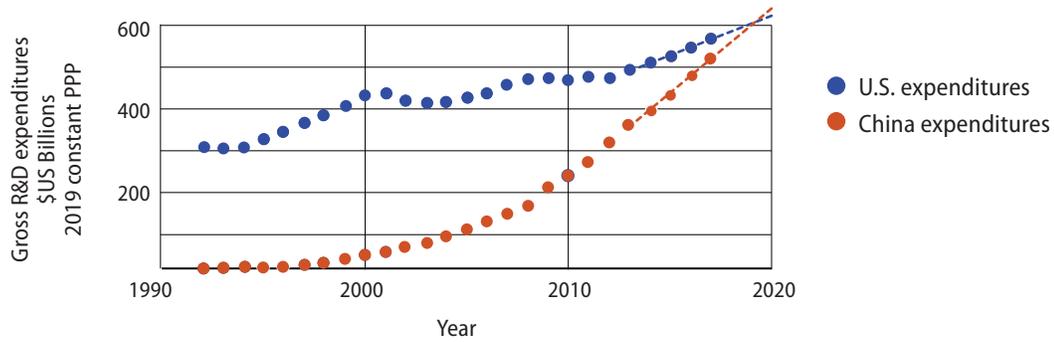

Fig. 2: US vs. China government R&D expenditures (reprinted from [2]).

The role of industry in addressing fundamental talent recruitment efforts is particularly important. A robust and vibrant domestic industrial base serves as a training ground for engineers and technicians who often have not considered the US government laboratories as career choices or lacked the experience in the field. Domestic industry has the potential to extend a much broader reach throughout the nation, which can provide on-the-job training and expand the relevant knowledge base for the high energy physics community. The growth of this qualified workforce can benefit industry and laboratories alike, and provide the potential to invigorate the aging workforce at the US laboratories, particularly in the engineering and technical roles. It is also not uncommon for industry to train up early career professionals who go on to accept roles within the domestic laboratory complex.

For those whom the national laboratory environment is not ideal, the private sector provides an alternative high-paced, diverse, and occasionally lucrative career path. These high-mix low-volume (HMLV) technologies are highly attractive for motivated technical personnel. The US accelerator industry often serves as a bridge between the broader accelerator community and other domestic industries. This is particularly important in manufacturing, where rapidly developing advanced processes and techniques have to be continuously adopted from other industrial sectors.

Finally, it may also be argued, that a growing industrial participation is a bellwether of the general growth and development in a particular technical field, and vice versa. For instance, the X-ray Free Electron Laser (XFEL) facilities represent a cornerstone of the modern international scientific infrastructure, benefiting a wide array of non-accelerator scientific and technical disciplines. XFEL technology was initially invented and implemented in the US, but never industrialized to the extent that was done in Europe and Asia. Now, only a decade later, we are witnessing a significant and growing quantitative gap of XFEL facility availability that places the US at a disadvantage, when compared to other regions (Fig. 3). Of course, industrial participation is not the only reason for such an outcome; however, if domestic industry played a more profound role it might be able to garner broader public support for the development of new light sources and other scientific facilities.

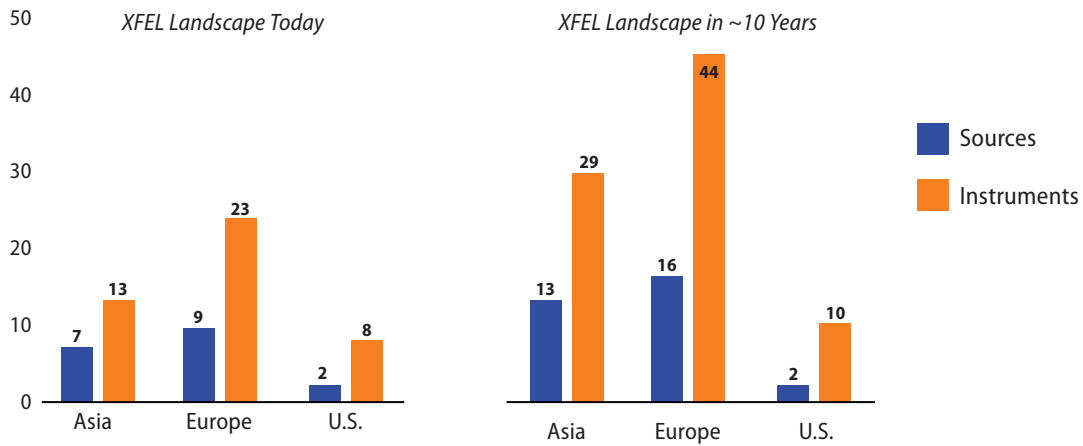

Fig. 3: Comparing numbers of independent XFEL sources (i.e., undulator beamlines) and instruments (i.e., physically separate, independent experimental stations) in the U.S., Asia, and Europe today (in 2021) vs. 10 years from now, based on announced projects (reprinted from [2]).

In summary, the existence of a dynamic, competent, and financially-viable domestic industry, (with healthy competition, of course), would in the long run benefit the high energy physics community and broader society in the following ways:

(1) reduce the cost and improve the quality of laboratory-based scientific programs,
(2) reduce US dependence on foreign, and not always friendly, sources of critical, sensitive, and high-impact technologies,
(3) improve economic and societal impact of the government investment in accelerator science, technology and facilities; and help to increase the volume and variety of such investments,
(4) help to widen and diversify the recruitment base of technical personnel, into labor pools hitherto inaccessible to the laboratories, due to geographical, social, or economic factors.

A competent industrial base, with strong professional ties to laboratories and universities, can play another important and traditional role of industry – to help shorten the cycle between new discoveries and their practical implementations to the benefits of the society. However, a detailed discussion of accelerator applications for the benefits of general public, is outside the scope of this paper, as we are focused on the domestic industry role as a vendor to DOE National Laboratories.

# 5 Challenges of selling products to the National Laboratories

## 5.1 Government regulations – why sell to the government?

Procurement of goods and services by the National Laboratories is governed by the Federal Acquisition Regulations, a thirty-seven chapter document whose first volume is over 900,000 words. For comparison, the Bible contains 780,000 words. The purpose of the regulations is to harmonize the government acquisition of goods and services, but its length, complexity and constant modifications places strain on the small companies that make up most of the domestic AT industry. Prudence dictates that the industrial supplier reviews relevant regulations thoroughly before submitting a bid or proposal. A simple widget that can be sold to another business for a small but profitable amount with little contractual overhead must be sold to the US government for the same price, but with much increased administrative effort, thereby reducing profitability. In many cases, it is *unprofitable* to sell to the government.

In such circumstances, one would expect that laboratories would actively look for vendors willing and able to play by these rules. Yet, the vendor outreach by National laboratory projects is practically non-existent. At conferences and trade shows, procurement officers do not walk the hall to see what is available, who vendors are, and what capabilities exist. Instead, procurement decisions are mostly reduced to proposal evaluations, using metrics designed to remove any human connection and trust between customers and vendors. As stated by Michael Pekeler (Research Instruments GmbH) at the Snowmass workshop held to discuss this topic [3], the high degree of mutual trust between his company and its customers has been one of the key components of company success and ability to take on risky projects in a cost-efficient manner. Without such trust, the contracting officer must consider every eventuality and protect the government from any known possibility of incurring unnecessary losses, or not getting the best deal achievable. As a result, in many cases, the overhead of responding to such requests for proposal (RFP) and engaging in subsequent contract negotiations becomes too excessive to motivate wide scale industrial participation.

## 5.2 Vendor Engagement

Industry *is* sometimes asked to participate in design decisions early in a project. Such early engagement is incredibly helpful, because the national laboratory can better understand the impact of certain decisions, and the industrial partner can better understand the soft requirements that are difficult to codify in an RFP. An excellent example is a workshop series devoted to beam position monitors (BPM) for the Advanced Photon Source upgrade (APS-U) at Argonne National Lab. Known as the APS-U BPM workshop series, it brought together APS-U engineers and physicists, world leaders, and industrial suppliers to discuss physics requirements of the APS upgrade's beam position monitors, *before* the engineering specifications were developed. While presentations by various attendees were given, it was the open and frank roundtable discussions, highlighting key constraints and goals, which directed the design iterations. The national laboratory understood what is cost-effective, where industrial R&D is needed, and what is potentially cost-prohibitive. On the industrial side, the full scope of the work became clear and allowed vendors to evaluate where further work would be needed and where costs could be reduced.

Unfortunately, such discussions and interactions are often limited during the design phase. Instead, National laboratory personnel provide a brief list of requirements and ask for formal quotes, with little desire to discuss what can be done, what will be more cost-effective, and what may need development. That is, a 'bare bones' system is quoted, much to the detriment of the project budget. Later, when the project is funded and a tender is fielded, the succinct original list of requirements is now dozens of pages long, fully designed by laboratory engineers for performance (not cost efficiency) while the available budget remains the same. To make an analogy, a four-door sedan is quoted early in the project, while the eventual requirement is a customized sports car.

There are, of course, intermediate levels of interaction between these two scenarios, but more often than not, they are minimal. It would be mutually beneficial, if laboratory engineers were encouraged to engage with industry throughout the procurement process.

## 5.3   Prototype development

Prototype development is a particularly difficult subject, because it is often expensive and is necessarily open-ended. Despite this reality, firm-fixed price contracts are all but demanded and technical requirements are rigid. There is no room for cost over-runs nor vendor response to customer-imposed scope creep. Likewise, there is little flexibility on performance. Project schedules, whether for an internal LDRD-funded effort or for a major construction project, are rigid. At least one leg of this triple constraint is destined to fail, with the full expectation by the laboratories that it be the cost, to be borne by the industrial supplier. Other contractual mechanisms are available, such as *cost plus fixed fee (CPFF)*, which allows more flexibility for the work to change and grow. Of course, it is still mutually beneficial for costs to be kept low, but there is a feedback mechanism to any rigid technical specifications and scope creep – the industrial supplier can be confident that the original scope will be maintained when cost increases have a direct impact on the project.

For industry, it is critical that the prototype procurement process should not be reduced to a minimalist build-to-print scenario. Industrial vendors of state-of-the-art accelerator components, microwave devices, and beam instrumentation need to be fully engaged from the start, and should be compensated for the full value of their products and services. This value is not limited to the cost of their manufacturing capabilities, but also reflects in-house expertise in design and engineering, as well as the development of special-purpose infrastructure.

To maximize the long-term success of the high energy physics community, it's important for laboratory procurement contracts to include relevant investments in industry, enabling full cost recovery for the required product or service (e.g., the novel BPMs require for the APS-U). This far-sighted approach would enable US companies to meet the needs of challenging projects, repeatedly over decade time scales, and subsequently to be competitive in bidding for comparable projects in other countries. What we are suggesting here is *against* the labs' short term interest, as it would force some of the DOE funds presently spent within the laboratories to be spent in industry; however, the long-term interests of entire community would be better served.

## 5.4   National Labs: partners, or competitors?

National laboratories are undoubtedly a center of US scientific talent. They have led the country in some of the largest scientific achievements and applications. They have disseminated countless discoveries and applications for the betterment of the US and its society. The recent contrasting push to make our National laboratories more self-funding and focus on commercialization is detrimental to this history of past success; how do our laboratories garner funding to do science?

To garner more funding, the laboratories are strongly focused on patents, licensing, paid technology transfers, strategic partnership projects (SPPs, formerly known as "Work for Others"), and other revenue streams. To support these revenue streams, laboratory designs and know-how are tightly controlled and treated as IP belonging to the lab's managing entity and are not accessible to the US public. Discussing in detail the associated merits or pitfalls of this is beyond the scope of this white paper, but there are some repercussions for industrial partnerships. Since foreign laboratories are willing to transfer knowledge to their local industrial base, the US scientific industry is at a disadvantage. In cases where laboratory technology is transferred to industry under a formal agreement, the process is prohibitively expensive and time consuming.

It is possible to have national laboratories perform work for the benefit of industry, and SPPs are an excellent way for this to happen. The main challenge here is that all work performed by the national laboratories is on a best-effort basis and comes with no guarantees. Such paid work is simply untenable for industry, where the survival of a commercial firm depends on delivering the product. Such best-effort "work per hour" arrangements are not easily available with industry as a provider. So, while transfer of knowledge is possible through an SPP, there is no assurance of success.

It is reasonable to assume that the next generation of AT products, which will be in demand 10 years from now, are being developed today in the national labs. Early industrial engagement with large development projects is the area where the tech transfer (TT) and knowledge transfer (KT) are most relevant. For this discussion, we refer to tech transfer as a process driven by the laboratories through their DOE funded TT offices, offering lab-developed IP to be licensed by industry. In contrast, we consider KT as direct vendor engagement with the expert laboratory personnel through collaborations, consulting arrangements, or most efficiently through laboratory supervision of prototype procurement contracts. In that sense, the TT process is more relevant for specific laboratory interests as a potential source of revenue stream from commercialization activities. In our view, KT is very different, because it leads to the development of institutional trust between laboratories and industry, driven by expert personnel. Hence, KT is the better path towards efficient technology industrialization. As stated by Giovanni Anelli at the Snowmass workshop on this topic [4], the KT metrics of CERN include: "maximize the technological and knowledge return to society, in particular through Member States industry".

Ideally, the DOE laboratories would purchase a prototype system from an industrial company and in the process of validation and qualification, use its superior expertise to teach that company to master the technology. In that process, the knowledge transfer is not just limited to the scientific staff, but is extended in a meaningful and lasting way to the engineers, manufacturing staff and technicians who often possess the key enabling know-how for industrial adaptation of novel technologies. If funding were to be earmarked for such activities in the future, it would foster meaningful long-term collaborations. Some smaller-scale examples of such engagements currently take place through the DOE SBIR/STTR program, serving as important examples that could be scaled up.

For government agencies, it does not make sense to invest in a company so it can produce a product that will not be in demand for another 20 years; however, careful planning of the national scientific infrastructure development includes identification of the technologies that will experience a sustainable demand. For these technologies, investment into vendor development is well justified. Thus, for the government to foster the development of the accelerator industrial base, policy makers and R&D leaders must analyze long term needs, identify critical gaps where sustainable industrial involvement is possible, and then strategically shift resource allocations from the laboratories to industry to cover those gaps.

# 6 Keeping the technological edge

Accelerator technology is, at its core, an advanced discipline. Not only do the results of the experimental efforts shape our understanding of the universe, but the knowledge gained along the way enable the next generation of projects and technologies to succeed. The boundaries are constantly being redefined and, in order to keep up with the pace of the international accelerator community needs, the technological capabilities of domestic industry has to grow also. In addition to talent recruitment programs, such innovation requires the funding and development of enabling infrastructure, whether it is computing resources, advanced manufacturing equipment and associated facilities or specialized electronics.

These forms of infrastructure represent high capital expenditures which are difficult to justify, given the challenging nature of working on laboratory projects (as discussed in the previous section). The SBIR/STTR program provides directed funding in chunks of approximately $1M over two years, and these relatively short-term directed research projects can effectively leverage such infrastructure but do not adequately justify its development.

Infrastructure investments represent a deflationary force on operations, as after the expenditure is committed, project execution efficiency can be improved. This efficiency comes in many forms depending on the investment, including but not limited to, reduced labor costs through automation, reduced scrap rates from more precisely controlled manufacturing, lower maintenance costs etc. In industry, these expenditures are either staged based on business growth and access to credit or accelerated through outside investments. As discussed further in a subsequent section, the latter mode of sourcing funding is not typically viable unless there is a spin off technology with a lucrative market that is independent of the AT market.

In other countries, the necessary investment in industry infrastructure is regularly accomplished through the use of government funding instruments or programs, which effectively subsidize operations. These high capital expenditures, which are essential to stand up, maintain and modernize manufacturing facilities, need not be included in the pricing of components. Given these factors, together with the historically lower wages in (for example) China, as compared to those in the United States, and the per unit cost of particle accelerator components from foreign companies can be difficult for domestic vendors to beat. These realities have forced the domestic accelerator industrial complex into a downward spiral, where technically demanding components, requiring highly-skilled labor, must be bid with low margins in order to compete against international vendors. Many US companies have experienced dire consequences, because low margins leave no room for error, while mistakes are inevitable when developing low volume, technically challenging products for an understandably demanding customer.

A positive approach would be to create channels for industry to engage deeper with the laboratories in the early stages of a project. Such engagement is not uncommon but is rife with difficulties at this time. The industry staff available for these discussions do not have significant billable hours to dedicate to this type of engagement and therefore much of this effort is pursued 'off the clock' or in the same vein as a loss-leader strategy. This limits the depth of effort and breadth of personnel industry can dedicate to these activities without significant internal investments and opportunity costs. If these activities can be funded directly from the DOE or industry carve outs placed in laboratory contracts, it would go a long way to promote early engagement with industry.

To be successful, the concept of quantifiable deliverables from laboratory procurement departments must be rethought in the above process. The goal must not be to provide any immediate benefit to the participating laboratory, but rather to prepare industry for potential upcoming projects. In these preparations, it is important to not just engage senior level scientists within industry but also foster mechanisms for engineering and technical staff to develop working relationships with their laboratory counterparts for critical knowledge transfer activities.

The intent of technology transfer programs within the laboratories is to assist in the above efforts. Unfortunately, the experience many industry professionals have had with tech transfer staff involves a focus on licensing revenue for specific inventions made at the labs. Although this is appropriate in certain circumstances, the vast majority of the value developed within the laboratories does not result in marketable products but rather in advancing the knowledge base of the community. If the primary avenue for industry to engage with laboratory expertise is to provide an offramp and revenue stream for lab-developed technologies, then we as a community have failed to see the big picture.

The forms of engagements presently fostered by TT programs are short lived and rarely if ever accompanied by critical pieces of infrastructure or equipment discussed above. We believe it would be mutually beneficial for the high energy physics community and for society at large, if the laboratories would provide matching funds in support of their technology licensing activities, in order to promote engagement with industry. We believe that IP-based revenue generation should not be a priority for the national labs. Those who have pursued tech transfer opportunities within the laboratories are sometimes met with legal and business staff who are incentivized to protect IP, further creating a barrier between the principal technologists from the laboratory and the company, thus further hampering rapid transition of technology from laboratory to industry.

One program which has proven to be a great resource for both the laboratories and industry is the SBIR/STTR program. Many burgeoning accelerator technology companies have been able to leverage this program and position themselves to provide significant value and overpressure valves for the myriad of projects ongoing within the labs. This unique and forward-looking program is unmatched within the world and often misunderstood. However, the true benefit of this program within the accelerator community is less about 'innovative research' as many of the innovations could have taken place within the laboratories and more about facilitating knowledge transfer through the engagement of laboratory 'end users' or collaborators. The most successful SBIR projects have motivated and engaged personnel on both sides freely sharing knowledge and resources towards a shared goal. This is less prevalent in typically more adversarial 'sales' activities and is a much more practical mechanism to transfer knowledge from the laboratories to industry. This program provides staged opportunities for a business to grow product lines, albeit at a slower pace than desirable, because the standard lifecycle for a project is three years with a two-year extension possible.

Some may argue that the SBIR program does not exist to fund businesses through the valley of death, a concept explained below but is intended to provide supplemental R&D funds for specific product development. This line of thinking assumes that the needs of the national laboratory accelerator complex represent a lucrative market that many businesses are eager to pursue and has a high enough volume of sales of any one specific product to justify the endeavor. However, it is more common that the requests for proposals that are specifically applicable to the accelerator complex are unique prototypes within a family of similar products. This also serves as a mechanism to partially cover non-recurring engineering costs that can assist in tackling the challenges of competition with foreign vendors or institutions.

Finally, another often overlooked cost driver that enables market penetration are the validation and quality assurance know-how and equipment required for a company to close the product life cycle loop. This enables companies to stand by their products with a deeper understanding of the end user requirements, rather than merely manufacturing to print. Although build-to-print type mechanisms have their place in procurement options, a business only capable of fulfilling such products will struggle to complete within the market at large with customers who may not have the necessary metrology in place to properly validate what was purchased. Encouraging open lines of communication with the technical staff engaged in these activities is a critical first step, which must then be followed up with mechanisms for industry to obtain the infrastructure required to carry out the activities. As major projects wind down, some of the equipment utilized to validate the myriad of procured products could be loaned to industry thus multiplying the impact of that government investment. Rather than being stuck with only being able to validate certain devices within the labs, industry could offer this service, additionally ensuring that industry personnel do not suffer from skill fade when the durations between projects drag into years, thus keeping industries' edge sharp and ready to compete. Accelerating this process will necessitate engaged guidance from laboratory personnel in setting up the validation capabilities and benchmarking the performance against the proven facilities at the labs.

# 7 The valley of death

The valley of death refers to the difficulty small startup companies have in transitioning their technology to market because of the difficulty of overcoming early negative cash flow. Amazingly, it is estimated that 90% of startups do not survive the first 3-5 years. Now, granted, this general figure is dominated by low-cost, non-manufacturing businesses that are not really relevant to an AT R&D supplier, but it does indicate the severity of the problem. The classic representation of this problem is illustrated in Fig. 4.

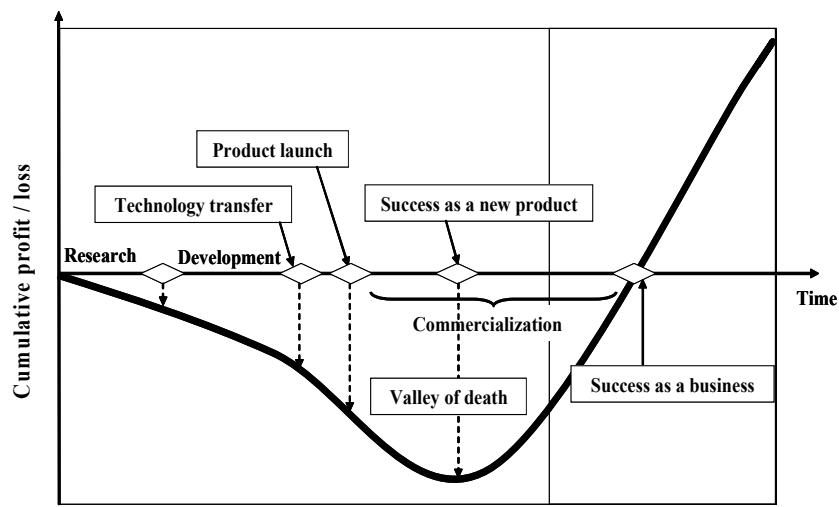

**Fig. 4: Overall process from Research to Commercialization (reprinted from [5]).**

The problem varies with the specific AT field that the new company is targeting. For a company with a new diagnostic, the problem is much less severe than for a company that intends to manufacture undulators, or even more so for SCRF cavities, where there is a huge initial infrastructure cost to be overcome in order to be competitive. This is upwards of $10M for SCRF cavities. In addition, the ultra-cleanliness required of SCRF cavities means that it is often undesirable to utilize the specialized equipment for material other than niobium. This is true of e-beam welders where one does not want, say copper residue, created inside the vacuum region of the welder. As a result, in the present SCRF market, it is well-nigh impossible to have enough work in a specific company to dedicate the expensive machinery to SCRF fabrication alone and an expensive piece of equipment with a low utilization factor provides little ROI. Angel or Venture Capital investors are not going to view a business model with such high initial startup costs and low equipment utilization factors very favorably when the downstream business return is both highly volatile and unspectacular, as is DOE AT R&D. It requires a hockey stick business plan to secure such investment, for example an extreme ultraviolet lithography (EUVL) free electron laser (FEL) or a novel high-energy ion implanter, so such companies must be able to support

themselves and generate profit for infrastructure investment from initiation. Generally speaking, that means they must have other core competencies that support a gradual growth in time into International competitiveness, with support from the DOE and National Laboratories. This is the MSU FRIB approach with Roark, a company that has a real profitable unrelated business and is interested in gradually expanding into AT R&D for SCRF cavities.

# 8 Case studies

It is important to begin this particular discussion by noting that there are many US companies that today successfully supply accelerator technology (AT) components to US DoE programs and to the Industrial marketplace. However, they generally fall into one of four specific categories. The first category is small business, SBIR (Small Business Innovative Research) houses, specializing in software (e.g. Tech-X, RadiaSoft), diagnostics (e.g. RadiaBeam, Euclid Technologies) and other mostly one-off products (e.g. Euclid Technologies for SCRF and other products). The second category is mid- to large-sized firms for whom DOE Accelerator Technology (AT) is a sideline that can be profitably exploited from time to time by their core competency, without impacting their core business (e.g. CPI and the LCLS-II HE fundamental power couplers). The third category is intermediate-sized companies producing AT components such as power supplies, magnets, high-vacuum products, electronics etc. for whom DoE Accelerator R&D projects are a secondary, but profitable, business. Finally, there are successful firms in established Industrial markets for lithography, medical X-ray, oncology and contraband detection accelerators (e.g. Axcelis, Varian, AS&E) that do not actively pursue DOE AT R&D projects because of the risk and volatility. There are also established companies that produce low-energy accelerators for Industrial applications like polymer crosslinking that have little relevance to DOE AT needs. What is missing are modest-sized US companies that fabricate undulators, SCRF cavities, precision magneto-optics and other specialized products. There is no US equivalent to Research Instruments GmbH.

In the past, there have been successful examples of North American companies that competed Internationally for specialized AT R&D business, but sadly they have gone out of business. The "Star Wars", and subsequently the Navy FEL program, spawned companies like STI Optronics and Advanced Energy Systems that in their heyday were at the International forefront of undulator and SCRF manufacturing sponsored by DOD funding. They were also key players in US DoE AT projects in the first two decades of the twenty first century. In Canada, TRIUMF engaged with Pavac during the same time period to help develop a Canadian SCRF capability but sadly, they also went out of business.

The domestic AT infrastructure problem remains today. The IsoDAR project is developing a cyclotron aimed at producing high fluxes of neutrinos for an underground experiment being deployed in Korea next to a three kiloton liquid scintillation detector. There is also interest in developing this cyclotron concept for isotope production, specifically for the alpha-emitting isotope Ac-225 that has shown remarkably good therapeutic effects. Four alpha particles are emitted from the same nucleus with huge destructive power if the isotope is placed in metastatic disease sites. The clinical experience is excellent and there is high demand, but the isotope is difficult to make though a promising means is spallation of protons above 50 MeV on natural thorium. Another way to produce it is via the (p,2n) reaction on Ra-226, but the target is highly radioactive and very hard to handle since the feed material is spent nuclear fuel. TRIUMF is using their large cyclotron to produce this isotope but the cyclotron is not always online. The issue for the IsoDAR group is where can they get their isotope cyclotrons built in the US where there is no obvious supplier. They are presently talking with IBA, in Belgium, who manufacture a good fraction of the world's isotope-producing cyclotrons but they are looking for and would like to find a domestic supplier.

## 8.1   SCRF Companies

There are two active US SCRF players at this time, Niowave and Roark. However, Niowave is focused on the Industrial medical isotope market with little present interest in DOE AT R&D, whilst Roark, though strongly engaged with FRIB, has not competed on the International stage to date. There was a third US company, Advanced Energy Systems (AES), which was a leading US International SCRF and NCRF accelerator supplier in the first two decades of the present century.

AES first became involved in SCRF on the Los Alamos National Laboratory (LANL) Accelerator Production of Tritium (APT) program, when the group was still part of Northrop Grumman. Despite the prevailing belief that no company could successfully deliver an SCRF cavity the first time around, AES delivered a 5-cell cavity that tested best of all cavities delivered under that program. In the meantime, AES had been spun out from Northrop Grumman in September 1998. Following the early APT success, AES became involved with Argonne National Laboratory (ANL), in time becoming their primary supplier for all formed and machined niobium components in their cavities built from about 2002 until 2016, when AES closed its doors. In parallel with the ANL work, AES was also working on a MW-Class elliptical cavity booster cryomodule for JLAB and the US Navy FEL. This, in time, led to the company becoming the key Industrial supplier for the Innovative Navy Prototype (INP) Free Electron Laser (FEL) SCRF accelerator to both Raytheon and Boeing. Along the way, AES developed SCRF guns with BNL, a MW-Class NCRF injector with LANL, and the NSLS-II 500MHz RF cavities.

In 2004, the International Linear Collider (ILC) program was initiated under the leadership of Fermilab and SLAC. AES was identified as a leading company in SCRF as by this time, they had delivered working cavities to customers. Fermilab was comfortable sending work to Europe to procure cavities, which they did in large numbers, but was also encouraging and enticing AES to get in the game. In 2005 AES was given a firm-fixed-price (FFP) contract for 4 ILC 9-cell cavities on which they took a large loss. It became clear AES needed to invest > $1.25M for in-house electron beam welding and clean room facilities. As all of this investment and improvement of the facilities and processes was taking place, it had a negative effect on the company's billing rates at a time when SCRF cavity pricing was relatively stable. It was at this same time that Fermilab also began courting new North American suppliers such as Niowave, Roark and PAVAC, in an effort to reduced cavity prices, thus putting additional pressure on AES profitability.

In the 2008-2010 time frame, AES was very much in contact with management at DESY with regard to the XFEL project. By this time, AES had delivered a series of top performing ILC cavities and had joined the club of "ILC Qualified" manufacturers. At that time, the club consisted of RI, Zanon, and AES. DESY was very interested in the possibility of having AES build some number of the 800 cavities for the XFEL. As time progressed however, EU officials realized that the only way they were going to get the number of cavities required within the time required, was to put substantial infrastructure into the companies. Furthermore, there would need to be a great deal of in-person support from Laboratory personnel at the contractors, particularly during the initial phases of the contract. This would be the case regardless of whether there were two or three cavity suppliers. Prior to issuing the requests for proposal, a decision was made that only EU suppliers would be allowed to bid to keep the monies local, an understandable decision that is not mimicked in the US.

In Spring 2012, the Navy abruptly terminated the INP FEL project and AES lost $24M in backlog overnight, a blow from which it never recovered. DOD funding had to a large extent enabled AES to continue pursuing non-profitable DoE AT SCRF projects. By 2014, AES focus had turned to LCLS-II. Unlike the EU officials on the XFEL, who evaluated the social impact of their project as well as the cost, US DOE and SLAC officials looked at their requirement for 280+ cavities, that were almost identical to the XFEL cavities, and concluded that they would buy from the same two European suppliers that were approaching the end of the XFEL production run. This was a windfall for RI and Zanon. The LCLS-II cavities needed one significant thing that the XFEL did not: the newly developed "Nitrogen Doping" process. This process, developed at US Laboratories and Universities, dramatically improves the cavity quality (Q) thereby significantly reducing cryogenic loads and the cost of the cryogenic system. This new technology, largely developed with American tax dollars, would have to be transferred abroad to the contractors building the LCLS-II cavities.

When the initial request for proposals came out for the LCLS-II cavities, there were certain minimum requirements given for the companies. One of these was that the company needed to have a "proven record" of delivering a similar number of cavities in the time required by the project. This requirement alone would have disqualified all prospective bidders except for the Europeans. AES successfully got that requirement removed but the handwriting was on the wall. So, the investment that the EU made in their companies for the XFEL truly paid off for them, with over $20M US dollars coming in from LCLS-II, and the cutting edge, US-developed, technology of nitrogen doping being transferred to RI and Zanon, making them the only companies in the world with that capability in-house.

AES continued to be involved in the LCLS-II and FRIB projects. AES cavities were used for the development of the nitrogen doping process and were the first cavities qualified for LCLS-II service. For FRIB, AES built the production run of 19 quarter wave resonators that are in the first cryomodules of the system. However, despite those successes, AES could never thereafter compete competitively in price in the EU or the USA with the successful EU SCRF manufacturers, and received no further LCLS-II or FRIB awards from those DoE programs, ostensibly because of cost. Attempts to create partial set asides for US manufacturers fell on deaf ears at DoE, SLAC and MSU. Curiously, stellar-performing AES prototype cavities populated early LCLS-II prototype cryomodules.

During this same period, AES was struggling with overruns on the fixed price NSLS-II RF cavities and the Fritz-Haber-Institut (FHI) der Max-Planck-Gesellshaft FEL. Though both of these programs were ultimately very successful they represented a large financial loss for AES. Coupled with the loss of Navy business, this forced AES to close its doors in 2016.

## 8.2 What lessons emerge from this AES case study?

One is a lack of consistency in contracting policy within the DOE Laboratories. Some have a very collaborative approach to contracting while others have an aggressively adversarial approach. Performance assessments for many buyers at the DOE Laboratories appear to be based upon achieving or exceeding financial savings goals. Rarely does it seem that buyers are evaluated on how well the product meets the technical requirements of the project. In many Institutions, the technical contract monitor appears to have very little authority to affect the course of the procurement up to and including evaluation of requests for out-of-scope budget adjustments. This was very much the case in the procurement by Brookhaven from AES of the cryomodules for the NSLS-II Light Source. Virtually every request AES made for budget adjustment based on scope creep was denied. Only after the project was in serious trouble and AES informed BNL that they would not be able to deliver the modules without budget relief, were many of the prior requests reviewed and ultimately approved.

Too many DOE contracts in the SCRF area have been issued as FFP delivery contracts when they were, in-fact, development contracts. This put great stress on US companies because they were aware that the bids were also going out to European companies who had been building similar cavities for DESY for many years and had been significantly facilitized for the XFEL. AES were forced to bid lower than they were comfortable with in order to win and then often took a large financial hit in order to deliver. The threat of buying from Europe was constantly used for leverage. At AES, it was not until the third contract for ILC cavities, after having delivered 10 units, that they managed to start making a profit on a procurement. This same process was repeated when Fermilab requested 650 MHz cavities for the PIP-II project, which was most definitely a development project for first-of-a-kind cavities but was issued as a FFP delivery contract. Despite Fermilab being a very collaborative player in these FFP contracts, there was little that could be commercially accomplished under that type of contract framework.

On the other hand, many of these contracts were issued as FFP but the scope was divided into in bite-sized pieces such that any risk involved in any particular step was not then compounded by many further steps also involving their own risk. Furthermore, the collaboration of the parties that were in partnership was extremely open and productive. This included the Argonne procurement group that was very tuned in to the work and ready to make contract modifications if required. It was by no means a blank check, but it was highly effective at responding to unexpected events while doing development work. Ken Shepard at ANL had a great deal to do with making sure this method of working was applied.

Finally, there is the taboo issue of US Laboratories competing with industry. In Europe, Laboratories like DESY do not build cavities. They generally do not even build prototypes. They will do process development, like hydroforming or spinning, namely, processes that are not yet part of normal manufacturing. When they need cavities, even prototypes, for testing, they go out to the companies to have them made. The Germans go to RI, the Italians go to Zanon, and the reminder of Europe typically goes to one or the other. Rarely, if ever, do they request quotes from outside the EU. In this way they maintain a flow of work through the companies even when large projects are not underway (see the financial importance of dedicated equipment utilization factors noted above). In the US it is quite different. Some National Laboratory routinely build prototype cavities in-house. They also build cavities for other Institutions such as universities. More than once, AES prepared bids to a US University for prototype cavities only to be told in the end "we decided to have" the National Lab "build them." It is not entirely clear that the Laboratory knew of the AES bids, but regardless they should never have undertaken to fabricate components for the University that they had to know could be furnished by US Industry. The ready availability of Staff and Facilities that are already covered in the overhead structure of the Laboratory, enables pricing that is impossible for a company to compete with.

Summarizing, this illustrates particular problems for small companies in the AT R&D field, where FFP contracts are a norm and yet the technology involved is state-of-the-art and constantly produces unexpected difficulties. Additionally, National Laboratories have a tendency to unwittingly introduce "scope creep" for which it is often difficult to negotiate appropriate compensation. Another common problem with small startup technology companies is that the management is often lacking in an adequate understanding of fundamental and essential business practices to ensure profitability, being somewhat blinded by the aura of their own technology. In the case of SCRF, all these problems are compounded by the fact that SCRF cavity fabrication is a not a real market. Programs are too spread out in time with large gaps and constantly sliding customer schedules to support a dedicated fabricator that does not have a main business in other products. Then there is always the ubiquitous gray boundary between what is proper National Laboratory R&D and what is competition with suppliers to muddy the waters.

### 8.3  Current market bottleneck in SCRF

AES fallout has happened in the most unfortunate time for the community, where the need for SCRF cavities have been growing ever since. The demand for custom SCRF cavities has increased significantly from most of the operating accelerator laboratories in addition to those under construction. In addition, there is a growing interest in SCRF industrial accelerators. Finally, there is a growing interest in SCRF cavities coming from many other fields of High Energy Physics in addition to accelerators, including:

- particle search (axion and dark photons) detectors,
- gravitation wave detectors, and
- quantum information systems (3D qubits, supported by OHEP).

In addition, SCRF cavities may also be of interest in material science.

All these research fields demand extended experimental work, and many trials and errors with various types of cavities. SCRF cavities are work horses for these experiments, but they are practically unavailable from the industry (18 months minimum wait time at the time of this writing). The EU vendors, RI and Zanon, are focused on mass production of SCRF cavities for large accelerator construction projects, and do not have a bandwidth for experimental structures, such as those needed at Fermilab for various ongoing research initiatives (Fig. 5).

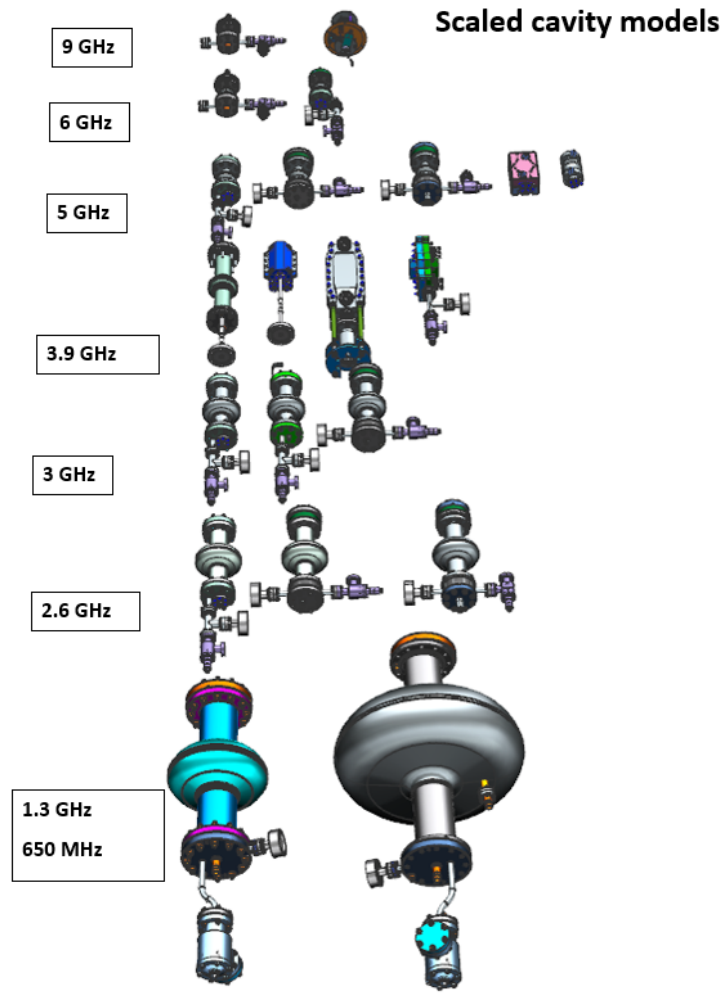

**Fig. 5: Menagerie of SCRF cavities required for various Fermilab experimental programs.**

This is an ongoing supply chain problem, and it will take a significant time, effort, and capital investments to fill the gap left by AES, for the "piece production" of experimental SCRF cavities, for the benefits of the scientific community. Painful as it is, such investment into the domestic vendors base is all but necessary to regain leadership and initiative in this growing applications space, most relevant to the HEP programs.

## 8.4   Undulator Companies

There is a similar unfortunate story to relay with respect to undulator fabricators. As noted above, STI Optronics (STIO) flourished under the Star Wars program building the 10m long permanent magnet NISUS hybrid undulator which was the amplifier in a MOPA with the THUNDER oscillator, which was also built by STIO. They won the White Sands Missile Range Ground-Based FEL (GBFEL), which award was worth over a $100M until it was cancelled. Note the parallel in backlog loss to AES. Between 1980 and 1995 STI designed and built eighteen undulators and wigglers for FEL and synchrotron radiation applications.

They then received a contract to build twenty four Advanced Photon Source (APS) undulator-A devices. After about 10 undulators on the original APS project the production rate for tuned, certified undulators was one every 2 weeks. However, delivery was delayed to once per month because APS could not absorb that production rate. Subsequently, eleven of these undulators were modified by STIO for the LEUTL SASE project at ANL. The STIO team was very experienced and professional. This was the high point for the company undulator business and profitability.

After the APS contract, STIO produced undulators for SRRC, retuned NISUS for BNL, fabricated the Brazilian L-frame LNLS wiggler, the JLAB FEL IR and UV undulators, two UCLA undulators, the NLCTA Echo-7 33mm and 55mm period undulators and the FHI IR undulator.

Just as with AES above, STIO was deeply involved in the NAVSEA FEL INP program and also suffered a backlog loss when the program was cancelled in 2012. During this same time period, STIO struggled because they could not compete effectively with the National Laboratories which usually kept the undulator design and testing functions in house. LCLS-I, LCLS-II, APSU, ALSU sought bids for machining or assembly only. It is difficult to be profitable on custom, one-of-a-kind, firm-fixed-price (FFP) devices. The non-recurring-engineering (NRE) alone can consume most of the profit for one-off devices. Then undulators are quite unforgiving of mistakes which you learn only at the end when they are being tested and when it is even more costly to fix. With no design work and no testing, the incentive to bid was low. LCLS-I wanted bids to assemble magnetic arrays per their instructions. They designed the undulator, procured the parts and would just ship pieces to the contractor for assembly. There was no participation in the design, no control of quality, and no testing, just assembly. STIO no bid. They hired a lobbyist and protested to no avail.

The Japanese at Spring8 heavily subsidized Sumitomo who built about 4 in-vacuum undulators (IVU) prototypes. Subsequently Laboratories wanted copies of the Spring8 design giving Sumitomo a significant edge. AVS|US, formerly AVS, which still exists, did build some IVUs but, for whatever reason, they are not a major competitive force in the International undulator business. KYMA and Danfysik were the principle European competitors for STIO. Curiously, Danfysik has recently removed undulator production from its catalogue of products offered but has not revealed why they chose to do this. One suspects they felt they can no longer be competitive in this niche market.

In 2016, STIO sold their undulator IP and infrastructure to Compass Engineering of Freemont, California, who saw an opportunity to enter the undulator business through the MARIE project. Unfortunately, there was a major change in management at Compass in 2018 that sought to return to the core competency and profitability of the company. Their interest in undulator fabrication evaporated overnight and their infrastructure purchases from STIO were sold to the Firitz-Haber-Institut (FHI) der Max-Plank-Gesellschaft in Berlin for the fabrication of an FHI and a FELIX undulator. Basically, 30+ years of experience was lost. Laboratories helped STIO when they allowed them to deliver complete, turnkey systems. Build to print sub-systems was never the STIO business model, but unfortunately that is all that seems to be on offer now from big projects.

## 8.5 Magnet Companies

Domestic magnet manufacturers capable of delivering particle accelerator grade magnets are few and far between with even fewer capable of managing a full product lifecycle internally. A complete lifecycle involves starting from a set of specifications, going into magnetic design, mechanical design, manufacturing of all critical components such as the yokes and coils, integrating these into a complete assembly and ends in full validation of the initial specifications. The magnet design requires expensive and specialized simulation codes and years of experience to run them correctly. The mechanical designs which account for manufacturing tolerance stack up and their consequences on the device quality can only be executed by trained and experienced staff. The often tight tolerance yet massive components create cost control challenges due to their manufacturing risks and the wide range of coil geometries desired require highly skilled technical staff. Highly skilled staff are also required to executed assembly and integration to laboratory quality standards. Finally, the validation of such devices, unless it is a build to print procurement, require specialized equipment and rarely mastered know how.

Considering the extremely low margins and high risks of undertaking these fixed price contracts provides little motivation for companies to attempt to enter this market. Additionally, the high barrier to entry discourages potentially interested companies from committing into this product line without significant financial support and guidance.

Buckley, which is a New Zealand company, dominates the present magnet supply market but their primary business is precision magnets for plentiful and highly-profitable high energy ion implanters for the semiconductor industry. This is yet another example where AT sales can be profitable for a company that has its main business in a different but related field.

## 8.6 Examples of successful SBIR-funded companies

Euclid Techlabs, LLC, which was formed in 2003, is a deep technology R&D company commercializing advanced designs, technologies, and materials for particle accelerators, electron microscopy, X-ray sources, and related industrial and medical applications. Since 2003, Euclid Techlabs has partnered with the national laboratories, including seven U.S. Department of Energy (DOE) labs, NIST, NIH, and Europe's CERN, as well as the electron microscope giant JEOL (Japan), to bring breakthrough technologies to market. Euclid has completed more than 30 Phase II SBIRs, and has been profiled as a commercialization "Success Story" on the DOE website [6], and in *2019 received the prestigious R&D 100 award* [7]. Euclid's 32 employees (16 PhDs), its portfolio of *seventeen patents*, and its strategic partnerships position it well in the particle accelerator, RF, X-ray, and microscopy markets.

Sometimes the most notable innovations in science come from linking traditionally separate or independent knowledge. Euclid has been a pioneer in the development of dielectric wakefield accelerators, and after extending its expertise to RF accelerators, Euclid has provided U.S. National laboratories with major innovative instruments and technologies over the years. However, it was the encounter of accelerator engineering and electron microscopy experts that launched Euclid's most significant commercialization achievement to date. Euclid's team realized that they could contribute their 20-year expertise in beam physics to solve a crucial limitation in electron microscopy and advance this already powerful technique to the forefront of in-situ characterization of nano- and atomic-scale processes. As a successful result of the DoE SBIR "Stroboscopic TEM Pulser" project, Euclid was awarded a Phase III contract for Transmission Electron Microscope (TEM) modification at NIST in 2016–2018, followed by a Chan-Zuckerberg award in 2021. The stroboscopic pulser has been installed in the leading TEM facility centers at BNL, NIST and JEOL. The modified next generation TEM pulser is being installed at CalTech and other locations in the US and Europe. The knowledge required to develop Euclid's electron buncher technology for the next generation ultrafast TEMs is strongly linked to technologies and products that Euclid developed through previous DOE SBIR grants, including the design and fabrication of turnkey accelerator systems and ultra-compact low energy accelerators, and a variety of accelerator components.

Unfortunately, transitioning from a prototype to a product is not always possible even with the most successful SBIR-funded projects due to the lack of follow-up funding mechanisms. An example is advanced materials such as ferroelectrics designed for fast high power tuners for SCRF accelerators. Previously, Euclid developed a robust new ferroelectric tuner for currently operating SCRF accelerators, providing the required level of microphonics compensation. This type of fast active tuning technology, applicable to SCRF cryomodules, is not currently available anywhere in the world. Previously, a new composite ferroelectric ceramic was developed by Euclid Techlabs, and a 400-MHz fast FRT prototype was developed and fabricated. Last year, this new type of fast tuner was tested at high power at CERN, and the technology was validated for microphonics compensation with a reduction factor of 14 and ~ µs response time. In fall 2021, Euclid Techlabs joined as an industrial collaborator with a new international program at CERN on the application of ferroelectric fast tuning technology for the Large Hadron Collider (LHC) transient detuning project.

Another illustrative SBIR story is RadiaBeam's "Aegis" RF gun, a high-performance electron beam driver for synchrotron light sources and other accelerators. RadiaBeam is another small business incubated and supported by the DOE SBIR program. RadiaBeam provides linac systems, accelerator components, instrumentation, and services for the research, medical, and industrial markets. The company currently has 54 full time employees and has been serving the US accelerator community and industrial markets since 2004.

The Aegis project started in 2016, when the Advanced Photon Source (APS) facility at the Argonne National Laboratory identified a need to upgrade the thermionic gun for the light source beam injection, as a critical component of the APS upgrade program (5 years down the line). The corresponding technical topic was introduced by APS to the DOE SBIR solicitation that year, and RadiaBeam applied and received a Fast Track DOE SBIR Award No. DE-SC0015191, to develop a thermionic gun compatible with the upgrade requirements of the APS. In the course of this project, RadiaBeam maintained continuous, open, and detail-oriented interactions with the APS scientists and engineers to receive real-time high-quality feedback on the design decisions from the intended laboratory customer. The prototype of the Aegis gun was industrially manufactured and tested at RadiaBeam, then delivered to APS. At APS, the injector was commissioned successfully [8], and has demonstrated continuous operation for 11 months, with the measured beam brightness improved by a factor of 2 over the previous versions of such gun.

Upon qualification, a Purchase Order for two additional thermionic gun units was placed by APS in 2021, an indication of an ultimate success for the SBIR-funded project. In addition to commercial success and benefits of supporting a high profile DOE program, Aegis project led to a spin-off experimental development of a high average power table-top THz sources, currently in progress in collaboration with UCLA. It has also led to recent industrial collaboration, to develop a system for non-destructive testing of power plant components.

The Aegis gun program at RadiaBeam is an excellent example of the DOE National Laboratory playing a proactive role in the SBIR program in support of the upcoming procurement needs. APS initiating the SBIR topic, stayed engaged with RadiaBeam throughout the project, tested and qualified the prototype, and purchased additional units upon qualification. That is exactly the type of long-term stewardship that the industry needs from DOE National Laboratories, to make itself useful, sustainable, and competitive.

# 9    The special case of software collaborations

For significant software development efforts, effective long-term collaboration between national laboratories, academia and industry will lead to important benefits for the entire HEP community. laboratories and universities will have access to better software with lower lifecycle costs. Companies will be strengthened by knowledge transfer from laboratories and universities. Computational scientists will be able to concentrate on core competencies, without spending time on UI design, ease of use, cloud computing, etc. Society will reap the benefits of better science, more innovation, and stronger businesses. State-of-the-art simulation codes will become readily available to students. Training time and associated costs will be reduced, as new team members will become productive more quickly. This will contribute to equity, diversity, and inclusion (EDI), as barriers to entry are removed for scientists in developing countries and for those at US institutions with less federal funding and no direct access to code developers.

## 9.1    Statement of the problem

The development and implementation of algorithms is a core competency of universities and research laboratories. When instantiated, the resulting codes often make use of a command line workflows which are error prone and difficult to reproduce. These workflows require excessive time and training to learn, involve multiple input and configuration files, execute on a high-performance server or cluster, necessitate post-processing with specialized software and additional visualization steps.

Professional software developers can make important contributions; however, they are expensive to hire and difficult to retain. Software sustainability and ease-of-use are very difficult to do well, but not especially interesting from the point of view of a computational physicist or computer scientist who needs to publish their work. Some excellent software developers and data scientists will be more easily hired, incentivized and retained in a small business environment, so close collaboration with industry can help to address career pipeline challenges with which the community is struggling.

Effective partnerships between industry and national laboratories, as well as between industry and universities, are necessary to maximize the productivity of available software development resources.

## 9.2 Particle accelerator codes

There are many world class particle accelerator design codes that are freely available to the community – a small subset includes MAD-X [9,10], elegant [11,12], Synergia [13,14], Zgoubi [15], OPAL [16], Warp [17-20] and JSPEC [21,22]. These codes offer wide ranging capabilities, with significant overlap, but each with uniquely important features. Each code varies in the difficulty required to compile and install, as well as the quality and quantity of user documentation.

Typically, users must learn a command-line workflow, which may involve script development and/or the understanding and editing of multiple input and configuration files. Typically, the codes are run in parallel on a Linux cluster or supercomputing center. The codes generate a variety of output and simulation results, sometimes in multiple files, using plain text and binary formats. Visualization and post-processing generally requires specialized software. The resulting workflows can be idiosyncratic, opaque and brittle. Some code development teams provide user support, but generally the important details of these complicated workflows are not available to scientists who do not have a good connection to expert users at major institutions.

Ease-of-use is important – a fact that is widely recognized by the development teams and by the community. However, it is not practical to develop a GUI for each separate code. The importance of code benchmarking and inter-comparison is also widely recognized, but there is not much incentive or reward for such efforts, so it is necessary to better facilitate the use of many codes together. A closely related problem is the difficulty of using multiple codes in sequence for beginning-to-end simulation of complex facilities.

Reproducibility and long-term sustainability are two important and related difficulties, which are not always adequately addressed. Simulations play an essential role in high-energy particle accelerator facilities over a period of decades, from pre-conceptual design, to final design, to commissioning and onward to a sequence of upgrades. It is essential that project scientists are able to reproduce past simulation results and to understand whether differences arise from improved modeling capabilities, changes in the design, or other factors. These concerns apply to many application codes throughout high energy physics.

## 9.3 Some requirements for success

In order to facilitate the necessary collaborations and to provide the entire community with confidence that the software will be widely available and adequately supported over decades, an open source license is required for the industry software and can be very helpful for the entire software ecosystem [23]. This imposes an open source business model on the corresponding businesses, at least with regard to this specific activity. The software design objectives must include seamless integration with legacy codes, low barrier to entry for new users, easily moving between GUI and command-line modes, cataloging of provenance to aid reproducibility, and simplified collaboration through multimodal sharing.

### 9.4 First example: Sirepo

Sirepo is an open source framework [24-28] for bringing scientific, engineering or educational software to the cloud, with a GUI that works in any modern browser on any computing device with sufficient screen size, including tablets. The Sirepo client is built on HTML5 technologies, including the JavaScript libraries Bootstrap and Angular. The D3.js library is used for 2D graphics, while VTK.js is used for 3D. The supported codes and dependencies are containerized via Docker, an open platform for distributed applications. RadiaSoft has developed open source software and expertise for building, deploying and executing scientific codes in Docker containers, and the corresponding images are publicly available. These containers are compatible with the Shifter containerization technology at the NERSC supercomputing center, which enables a Sirepo server to automatically launch jobs at NERSC.

A free Sirepo scientific gateway is available to the particle accelerator community [26], providing a broad selection of supported codes. The accelerator tracking codes include MAD-X, elegant, Synergia, OPAL and Zgoubi. Presently under development, the MAD-X sequence file format will be used as a common format to enable rapid code benchmarking and sequential use of multiple codes. The loosely coupled cloud-based architecture of Sirepo enables coupling with other sophisticated software and systems. This is another reason for the enterprise approach. At NSLS-II for example, the DAMA group is integrating Sirepo/SRW with their BlueSky [29,30] software for experimental control and data management.

Sirepo has been designed to transcend the limitations that discourage many scientists from working with GUI-driven applications. Sirepo can import the necessary input, data, or configuration files for the codes that it supports, so experts can quickly transfer their simulation results to a GUI user. Likewise, the GUI can export a zip file with everything needed to run the identical simulation from the command line.

### 9.5 Second example: Computational Model Builder

Computational Model Builder (CMB) is an open source platform with integrated software tools designed to integrate all processes involved in the life cycle of numerical simulation [31]. CMB is developed with a modular, flexible architecture that has been customized for several different scientific fields including hydrology, computational fluid dynamics, and multiphysics casting simulation. In high energy physics, CMB provides a graphical user interface for the ACE3P accelerator modeling codes. In every CMB application, the primary goal is to simplify end-user effort and reduce the manual overhead often taken up by workflow and data management activities associated with simulation-based design and analysis.

To prepare simulation inputs, CMB provides form-style fields for entering data combined with selection and highlighting of modeling geometry in 3D views. The input fields are automatically generated from XML template files that describe and organize the keywords making up the simulation code input specification. The UI includes syntax checking to reduce the likelihood of entering invalid data. At the backend, Python scripts are used to write the simulation input files based on the user-entered data. CMB allows new applications to be developed with less effort than custom UI software.

For simulation job execution, CMB relies on the Girder data management platform [32] as a middle-tier server connected between the desktop user and remote HPC or cloud-based systems. When users submit jobs from the CMB desktop, execution status is tracked continuously by Girder and reported back to the desktop. For ACE3P simulation, a Girder server has been deployed on the NERSC Spin platform for submitting and tracking simulation jobs. This system is in the process of being updated with additional resource-location services so that simulation results can be more easily traced back to their source data.

Because the CMB platform is built on ParaView [33], it provides the full set of ParaView postprocessing and visualization features. This includes remote visualization of simulation results and in situ visualization of interim results during execution. CMB and ACE3P are currently being updated to support these features so they can be seamlessly accessed from the CMB user interface, again reducing the effort required by scientific researchers. As with all aspects of the CMB design, the overall goal is to adapt to the needs of the simulation user, instead of requiring the user to adapt to the available computing environment and software tools. For applications that integrate geometry modeling, CMB also includes a geometry module for operations such as model creation, model modification, and discretization using external meshing technologies.

# 10 Conclusions

In this white paper we established that industry plays an important role in the scientific community and society in general, and yet the US domestic industry serving the needs of the DOE accelerator facilities has been struggling to achieve prominence. The AES and STI Optronics pioneered SCRF and undulator technologies, respectively, but could not sustain their business models without long term programmatic support from the customers. DOE laboratories have been indifferent to the plight of these companies, resulting in their eventual downfall.

In both scenarios, the companies initially successfully acquired expertise, equipment, experience, and motivated customers base, all at a great cost and through decade-long efforts. Yet at the end they only proved that in such high-mix low-volume niche markets, when only serving the accelerator equipment needs of the DOE labs, a business cannot sustain itself without directed long term investment by the funding agencies, as it is practiced in Europe and Asia. The collapse of these companies created supply chains gaps, from which to this day our field has never fully recovered.

There are number of currently active SBIR-funded companies that have achieved considerable capabilities and expertise (i.e., RadiaBeam, Euclid, RadiaSoft) and are actively involved in many activities within the DOE accelerator complex. However, the same lack of long term directed funding beyond the SBIR prototype stage creates a headwind for these companies, preventing them from expanding their role in the US National Laboratories ecosystem.

If the goal is to nurture and sustain a vibrant and competitive accelerator technology domestic industrial base in the US, some regulatory changes need to take place. First of all, the DOE funding mechanisms that already exist, could be better utilized towards this goal. For instance, SBIR program topics could be better aligned with the future procurement programs in the labs. In an ideal scenario, once a small business completes the prototype under an SBIR award, the National Laboratory promoting that SBIR topic, would qualify the prototype, and issue a follow up purchase order for multiple production units.

Besides the SBIR program, there are many other programs that support commercialization activities at the labs. Recent decades saw a proliferation of National Laboratories based commercialization centers built around the technology transfer activities. Yet, few can report successes and, as was discussed in Ref. [4], the idea of technology transfer that can help funding the laboratories in theory sounds great, but in practice is often futile. Accelerator technology facilities differ from the university IT or biotech incubators: they work with heavy equipment which takes many years to develop, and a lot of build-up infrastructure to produce, with a relatively small and not scalable number of customers. Such line of business is highly unattractive to venture capitalists and "shark tank" investors, and it should be unattractive to the laboratories as well. We believe it would be more beneficial to deemphasize technology transfer as a means of supporting the labs, and emphasize knowledge transfer as a means of supporting motivated businesses to expand capabilities of interest to the DOE programs. Laboratories should welcome an industry interest in use of their expert consultants, specialized equipment, and IP, to develop cutting edge and economically viable commercial solutions that eventually benefit the accelerator community as a whole.

We also discussed a need to simplify some of the laboratory procurement practices, and likewise explore various creative ways for industry and laboratories to collaborate on the prototype developments that would minimize the risks and maximize return to both sides. The accelerator community should also promote programs that facilitate direct and open communication channels between laboratory engineering and technical staff with their industrial counterparts (there are many conferences for scientists to attend and share their experiences, but not so many venues are available to technicians and engineers whose skills are essential and irreplicable in our field).

These and many other steps could improve the quality and outcome of industry participation in the US DOE accelerator laboratories complex. Without proactive joint programs, and without having a seat at the table, it would be difficult for the US industry to achieve and sustain the prominence and excellence required to compete internationally.

Finally, we recommend that DOE establish a method to identify key technologies that will be needed in a decade time frame and create new channels of direct funding of the industry to develop infrastructure and capacity to meet such needs. It is equally important to be able to help sustaining the companies that have already achieved critical capabilities, but are not able to sustain them without a minimum volume of recurrent orders.

# 11 Acknowledgement

The authors would like to thank Giovanni Anelli, Eric Colby, Vitaly Yakimenko, Michael Pekeler, Timur Shaftan, Ed Bonnema, Salime Boucher, and Emilio Nanni for their contributions to the "Industrial Base Development for HEP needs" mini-workshop and many follow-up useful discussions.

Contributors to Software section include: R. O'Bara, J. Tourtellott (Kitware), R. Nagler, P. Moeller, D.T. Abell, E. Carlin, S. Coleman, N.M. Cook, J. Edelen, C.C. Hall, M.V. Keilman, B. Nash, I. Pogorelov (RadiaSoft), S. Baturin (NIU), A.F. Habib, T. Heinemann, B. Hidding, P. Scherkl (U. Strathclyde), A. Huebl, R. Lehe, J.-L. Vay (LBNL), C.-K. Ng (SLAC), C.S. Park (Korea University), P. Piot (NIU, ANL), A. Sauers (Fermilab)

The authors are also thankful to Daria Wang for the administrative and organizational help in preparation of this white paper.